\documentclass{aastex}
\usepackage{emulateapj5}

\newif\ifAMStwofonts

\AMStwofontstrue

\def\ga{\mathrel{\hbox{\rlap{\hbox{\lower4pt\hbox{$\sim$}}}\hbox{$>$}}}}
\def\la{\mathrel{\hbox{\rlap{\hbox{\lower4pt\hbox{$\sim$}}}\hbox{$<$}}}}

\shorttitle{The thermally-unstable warm neutral medium}
\submitted{Accepted by ApJ}

\shortauthors{}

\begin{document}

\title{The thermally-unstable warm neutral medium: key for modeling the ISM}

\author{Ayesha Begum$^1$, Sne\v{z}ana Stanimirovi\'{c}$^1$, W. M. Goss$^2$, Carl Heiles$^3$, 
Anthony S. Pavkovich$^1$ \& Patrick Hennebelle$^4$}

\affil{
$^1$University of Wisconsin, Madison, 475 N Charter St, Madison, WI 53706,
$^2$National Radio Astronomy Observatory, P.O. Box O, 1003 Lopezville, Socorro, NM 87801 
$^3$Radio Astronomy Lab, UC Berkeley, 601 Campbell Hall, Berkeley, CA 94720,
$^4$Laboratoire de radioastronomie, UMR 8112 du CNRS, 24 rue Lhomond, 75231 Paris, France
}

\email{begum@astro.wisc.edu, sstanimi@astro.wisc.edu}

\begin{abstract}
We present 21-cm absorption measurements towards 12 radio continuum 
sources with previously identified thermally-unstable warm neutral medium (WNM).
These observations were obtained with the Expanded Very Large Array (EVLA)
and were complemented with the HI emission spectra obtained with the Arecibo Observatory.
Out of 12 sources, HI absorption was detected along 5 lines of sight (seven
new absorption features in total), resulting 
in a detection rate of $\sim$ 42 \%. 
While our observations are sensitive to the WNM with a spin temperature $T_s<3000$ K,
we detected only two wide absorption lines with $T_s=400-900$ K.
These temperatures lie above the range allowed for the cold neutral medium (CNM)
by the thermal equilbrium models and signify the thermally unstable WNM.
Several absorption features have an optical
depth of only a few $\times 10^{-3}$. While this is close
or lower than what is theoretically expected for the CNM, 
we show that these weak lines are important for constraining the fraction of the
thermally unstable WNM.
Our observations demonstrate
that, for the first time, high bandpass stability
can be achieved with the VLA, allowing detection of absorption lines with a peak
optical depth of $\sim 10^{-3}$.
\end{abstract}



\section{Introduction}

The diffuse neutral interstellar medium (ISM) is known to exist in two flavors: 
the cold neutral medium (CNM) and the warm neutral medium (WNM). 
Traditionally, the CNM and WNM are understood as being
two thermal equilibrium states of the neutral ISM
with  temperature of $40 - 200 $ K and $\sim 8000$ K \citep{Field69,MO77,Wolfire03}.
In addition, for the CNM the spin temperature, ${\rm T_s \sim T_k}$ 
is expected from theoretical considerations.
The situation is more complicated in the case of the WNM and $T_s$ could be lower
than $T_k$ \citep{liszt}.
Observationally, the CNM is easily studied with the 21-cm line absorption.
While the CNM properties have been measured extensively,  only three direct 
measurements of the WNM spin temperature exist to date \citep{kanekar,dwaraka02,carilli98}.
The main reason for this observational
paucity is the low optical depth of the WNM, $\tau \la 10^{-3}$, which creates a need for 
very sensitive radio instruments. Properties of the WNM are traditionally indirectly
inferred only through HI emission line profiles. Out of all ISM phases, the WNM is the
least understood; yet it seems to hold the key for constraining ISM models and the formation
of CNM clouds.





The theoretical and numerical models of the ISM take into account heating and cooling
processes and try to reproduce the observed physical properties of the CNM and the
WNM.  These properties include temperature distribution, 
column density probability density functions (PDFs),
the gas fraction in each phase, and even profiles of emission and absorption spectra.
\cite{MO77}, and similarly \cite{Wolfire03}, predict that most of the WNM, which in their
model envelopes the CNM, should be in thermal equilibrium and comprise only
about 4\% of the total column density. However, observers have traditionally
found a much higher fraction for the WNM.

In addition, a significant fraction of the WNM could be thermally unstable.
Using a statistically meaningful sample and the Arecibo radio telescope,
Heiles \& Troland (2003; HT03) found that 48\% of the WNM in their survey 
was in the thermally unstable phase, with kinetic temperatures in the range of 
500 $-$ 5000 K. The HT03 results
question the meaning of a two-phase neutral medium and
have triggered substantial numerical studies. 
Numerical models of the ISM which include turbulence 
easily find thermally unstable gas and lead to the conclusion that dynamical processes 
overshadow the microscopic ones in determining gas temperature. 
Macroscopical processes are hard to calculate theoretically 
because they depend on many forms of energy input (HT03).
As a result, the predicted column density 
fraction of the thermally unstable gas varies wildly; from 
very large values strongly dependent on the SN rate, as in \cite{maclow05},
to $\sim30$\% in \cite{ham}.

\begin{table*}
\caption{EVLA Observations }
\centering
\label{tab:obs}
\begin{tabular}{lccccccc}
\hline
\hline
Source  & RA(J2000) & Dec (J2000) & S$_{1420}$ & rms & $\tau_{rms}$ & Synthesized beam & Time \\
        &  (Deg)      &  (Deg)        &  (Jy) & (mJy) &           &  (arcsec$^2$)    & (hrs) \\
\hline
J0022+002 & 00:22:25.4 & +00:14:56 &  3.05 & 4.4 & 0.0014 &  14.0 $\times$ 10.0& 5.0\\
J0025-260 & 00:25:49.2 & $-$26:02:12 &  8.81 & 7.5 & 0.0009 & 17.0 $\times$ 12.0 & 1.0 \\
P0347+05$^*$ & 03:49:46.7  & +05:51:38 & 1.96 & 4.7 & 0.0024 &  14.0 $\times$ 9.0 & 5.0\\
3C234$^*$ & 10:01:49.0 & +28:47:12 & 2.12 & 8.0 & 0.0038 & 13.0 $\times$ 9.0  & 2.0\\
3C236 & 10:06:01.8 & +34:54:12 &  3.22 &  4.8 & 0.0015 & 13.0 $\times$ 9.0 & 5.0\\
3C263.1 & 11:46:19.9 & +22:04:10 & 3.23 & 3.8 & 0.0011 & 13.0 $\times$9.0 & 5.0\\
3C273$^*$ & 12:29:06.4 & +02:03:05 & 32.7 & 10.0 & 0.0003 &14.0$\times$9.0 &1.0\\
3C287& 13:30:37.7 & +25:09:10 & 6.69 & 6.0 & 0.0009 & 13.0$\times$9.0 & 1.0\\
3C286 & 13:31:08.3 & +30:30:33 & 16.0 & 3.2 & 0.0002 &  13.0 $\times$ 12.0 & 10.0\\
J1613+324 & 16:13:41.1 & +34:12:47 & 4.40 & 5.0 & 0.0011 & 12.5 $\times$ 9.0&2.0\\
J1634+627 & 16:34:33.8 & +62:45:35 & 5.14 & 5.8 & 0.0011 &  13.0 $\times$ 10.0 &2.0\\
J1635+381 & 16:35:15.5 & +38:08:04 &  2.52 & 4.0 & 0.0016 &  13.4 $\times$ 8.6 &5.0\\
\hline
\end{tabular}
\vskip 0.1 in
${}$~$^*$~The continuum source is resolved with the EVLA observations. \\
\end{table*}

However, the estimated fraction of the thermally unstable WNM in HT03 could be
significantly biased by the low sensitivity of this survey. In many of the observed directions 
(16 out of 79 in this survey) the CNM was not detected at all. In addition, 
there were many cases where relatively narrow WNM components (based on Gaussian fits)
were found without corresponding CNM features.
These WNM components often implied kinetic temperatures of 2000-3000 K and are
typical representatives of the thermally unstable phase.
As discussed in HT03, one possibility is that these lines of sight trace 
special conditions where
the CNM was destroyed in some way (SNe etc). However, the integration time of only
15 min per source resulted in a typical optical depth sensitivity of $\sim 10^{-2}$.
\cite{sh05} showed that with longer Arecibo integrations, weak 
absorption lines, with $\Delta \tau \sim 10^{-3}$, are easily detected.
These weak absorption lines play an important role in the fitting of HI
emission spectra and the derivation of the spin 
temperature. If commonly found in observations, the 
weak absorption lines would result in a decrease of the estimated fraction of 
the thermally unstable WNM.

In this paper we report the results of deep HI absorption observations with the
Expanded VLA (EVLA) in the directions of thermally-unstable WNM. The main goal of this 
project was to test the feasibility of sensitive HI absorption measurements with a peak optical 
depth $\tau \sim$ a few $\times 10^{-3}$ with the EVLA.
In two independent pilot projects carried out with the VLA in 2006 \& 2009, 
we observed a sample of 12 continuum sources. The sources were chosen from the previous 
emission/absorption surveys of HT03 and \cite{mohan}. 
We specifically focused on lines of sight where previous observations
indicated the existence 
of the thermally-unstable WNM with a kinetic temperature $<$ 5000 K. 
All continuum sources in our sample are at high Galactic latitudes i.e. $|b|>40^\circ$, to
avoid complicated CNM profiles.

The structure of this paper is organized in the following way. Section~\ref{s:obs}
summarizes the EVLA observations and the data reduction. Our results are presented
in Section~\ref{s:results} and discussed in Section~\ref{s:discuss}. 
We summarize our findings in Section 5.

\section{Observations and Data Processing}
\label{s:obs}

\begin{figure*}
\epsscale{2.0}
\plottwo{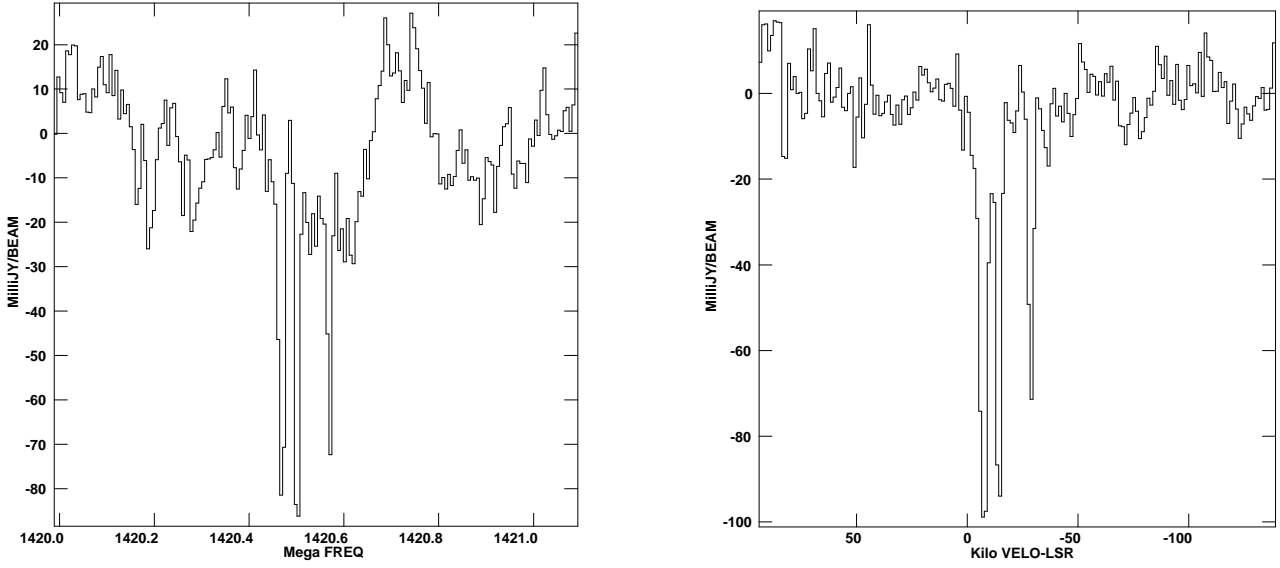}{evla.ps}
\caption{\label{fig:bandpass}
The HI absorption spectrum in the direction of 3C286 obtained with VLA antennas only (left),
and the EVLA antennas only (right). The bandpass stability is largely
improved with the use of EVLA antennas and
allowed a clear detection of spectral lines with very low optical depth.
The deepest absorption feature in the right-hand
figure corresponds to  $\tau \sim 6 \times 10^{-3}$.
}
\end{figure*}

\begin{table*}
\caption{The Gaussian parameters from the fit of the EVLA absorption spectra}
\centering
\label{tab:cnm}
\begin{tabular}{lccccccc}
\hline
\hline
Source  &  $\tau_{\rm peak}$   &          V$_{\rm LSR}$ &     FWHM &     $T_{\rm spin}$  &        T$_{k,max}$ &     N(HI)$_{\rm CNM}$  &          N(HI)$^{\rm Upper}_{\rm CNM}$ \\
       &             &         (kms$^{-1}$)       & (kms$^{-1}$) &  (K)      & (K)    & (10$^{20}$ cm$^{-2})$ &(10$^{20}$ cm$^{-2}$)\\
\hline
J0022+002  &   0.0074 $\pm$ 0.0011 &    $1.58$  &   7.92  &  421 $\pm$ 46  & 1358  &  0.477 $\pm$ 0.158   &  1.54 \\
J0022+002  &   0.0132 $\pm$ 0.0018 &   $-5.51$  &   4.18  &  137 $\pm$ 33  &  377  &  0.147 $\pm$ 0.048   &  0.40 \\
P0347+05   &  0.9564 $\pm$ 0.0199  &    $8.09$  &   3.18   &  61 $\pm$ 5  &   219  &  3.609 $\pm$ 0.113 &   12.97 \\
P0347+05   &  0.2972 $\pm$ 0.0168  &   $9.75$   &  7.31   & 148 $\pm$ 3   & 1155  &  6.266 $\pm$ 0.398  &  48.89 \\
P0347+05   &  0.0459 $\pm$ 0.0046  &  $-0.47$   &  4.05   & 175 $\pm$ 12  &  355  &  0.632 $\pm$ 0.112  &   1.29 \\
3C263.1 &  0.0252 $\pm$ 0.0024 &   $-54.12$ &    2.22  &   25 $\pm$ 10 &    107 &   0.027 $\pm$ 0.004  &   0.12 \\
3C273   &  0.0177 $\pm$ 0.0016  &  $-6.28$   &  2.59   &  71 $\pm$ 5   &  146  &  0.063 $\pm$ 0.009  &   0.13 \\
3C273   &  0.0036 $\pm$ 0.0017  &  $-5.89$   &  7.77   & 910 $\pm$ 19  & 1307  &  0.516 $\pm$ 0.258  &   0.74 \\
3C286  &   0.0046 $\pm$ 0.0004 &  $-28.79$  &   2.66  &   95 $\pm$ 16  &  153  &  0.024 $\pm$ 0.005 &   0.04 \\
3C286  &   0.0067 $\pm$ 0.0004 &  $-14.26$  &   2.58  &   39 $\pm$ 12  &  144  &  0.013 $\pm$ 0.004  &   0.05 \\
3C286  &   0.0064 $\pm$ 0.0003 &   $-7.39$  &   4.49  &   26 $\pm$ 18  &  437  &  0.015 $\pm$ 0.010  &   0.25 \\
\hline
\hline
\end{tabular}
\end{table*}

We observed 3C286 in April 2008
with the VLA during the configuration move CnB$\rightarrow$C and using 14 EVLA antennas. 
HI absorption observations of additional 11 sources were carried out
in May 2009 during the B$\rightarrow$C configuration move and using 20 EVLA antennas.

In Table~\ref{tab:obs} we list coordinates and the 21-cm flux density of all observed sources, 
as well as a few observational details. 
Column (1) gives the source name, columns (2) and (3)  
the source RA and Dec in J2000 coordinates, column (4) the flux density at 21 cm
determined from our EVLA observations, column (5) the rms noise in each line channel 
in mJy, column (6) the rms noise per channel in the optical depth units, 
column (7) the synthesized beam
size, and column (8) the total on-source integration time.

For all observations we used a bandwidth of 1.56 MHz, centered at the 
frequency of the 21-cm HI line, 
with 256 spectral channels. The resulting velocity resolution is $1.56$ km s$^{-1}$. 
For absolute flux density calibration we observed sources 3C286 and 3C48. 
Since most of our the sources are bright, amplitude and phase calibrations of 
the antennas were carried out through self-calibration. The bandpass calibration was 
obtained through frequency switching on the source itself. 
For this purpose, each 15 minute``on-line" 
scan was sandwiched between two 7 minute ``off-line" scans.
The off-line scans were offset in frequency from the on-line scans by 1.56 MHz and 
were not affected by contamination due to Galactic HI.
This observing strategy provided accurate bandpasses. To test the quality of
our bandpasses, we used one of the ``off-line" scans as a bandpass for 
the other ``off-line" scan. This resulted in a flat spectrum with
a rms noise of $2 \times 10^{-3}$ for weak sources such as 3C263.1,
and $\sim 10^{-4}$ for stronger sources.

The visibility data was analyzed using the Astronomical Image Processing System (AIPS). 
For each source, line-free channels were averaged to construct a single continuum
channel. The amplitude and phase antenna solutions were obtained through normal
and self calibration of this continuum channel and then transferred to the spectral line
visibility data. The calibrated and bandpass corrected spectral line data
was used for further analysis.  The continuum was removed using the AIPS 
task UVLSF by performing a polynomial fit to the visibilities in the 
line-free channels and subtracting it from the source visibilities in all the  
spectral channels. For most sources a  first or a second-order fit was  used 
to remove the continuum. In the case of two sources,
3C286 and J1635+381, a 3rd-order polynomial
was required to obtain a flat baseline.
Before continuum subtraction the spectrum from each source was carefully inspected
to select line-free channels. The continuum 
subtracted visibilities were used to make spectral cubes of the HI absorption 
in the direction of each source.  
The spectral cubes were CLEANed using the AIPS task IMAGR.

The bandpass stability was greatly improved in the case of the upgraded VLA antennas.
Figure~\ref{fig:bandpass} shows the HI absorption spectrum in the direction of 3C286,
obtained using the VLA antennas only (left) and the EVLA antennas only (right). 
The improved bandpass stability  allowed a clear detection of spectral
lines with a low optical depth of a few  $\times 10^{-3}$. 
Hence, for all sources, only the EVLA antennas were used
to produce the final HI absorption spectra.

In order to estimate the spin temperature of the detected HI absorption components, the 
EVLA absorption profiles were complemented by the Arecibo HI emission 
profiles with an angular resolution of 3.5 arcmin (HT03). 
The emission spectra for all sources except 3C286 and J0022+002 were obtained from HT03. 
Since  the source J0022+002 was not observed by HT03, 
its emission spectrum  was kindly provided 
to us by Lou Nigra. This spectrum was also obtained  with the 
Arecibo telescope using exactly 
the same observing technique as in HT03. 
In the case of 3C286, the HT03 emission spectrum was very noisy
and we used the Arecibo emission spectrum from \cite{sh05}.

\section{Results}

\label{s:results}

\begin{table*}
\caption{The Gaussian parameters from the fit of the emission spectra}
\centering
\label{tab:wnm}
\begin{tabular}{lccc}
\hline
\hline
Source  &     V$_{\rm LSR}$ (kms$^{-1}$) &   FWHM (kms$^{-1}$)  &         T$_{k,max}$ (10$^3$) K\\
\hline
J0022+002   &   $-$2.28   &    22.39  &     108.5\\
J0022+002   &   $-$4.66   &    8.41   &    1.5\\
P0347+05   &    4.95   &    17.33  &     6.5\\
3C263.1 &   $-$44.22  &     37.08 &      29.8\\
3C263.1 &   $-$39.09  &     8.37  &     1.5\\
3C263.1 &   $-7$.04   &    20.04  &     8.7\\
3C273   &    24.92  &     23.71 &      12.1\\
3C273   &   $-$15.56  &     35.49 &      27.2\\
3C286   &   $-$5.92   &    50.19 &      54.5\\
3C286   &   $-$6.08   &    11.35  &      2.8\\
3C286   &   $-$26.45  &     22.15 &      10.6\\
\hline
\hline
\end{tabular}
\end{table*}

\begin{figure*}
\epsscale{2.0}
\plottwo{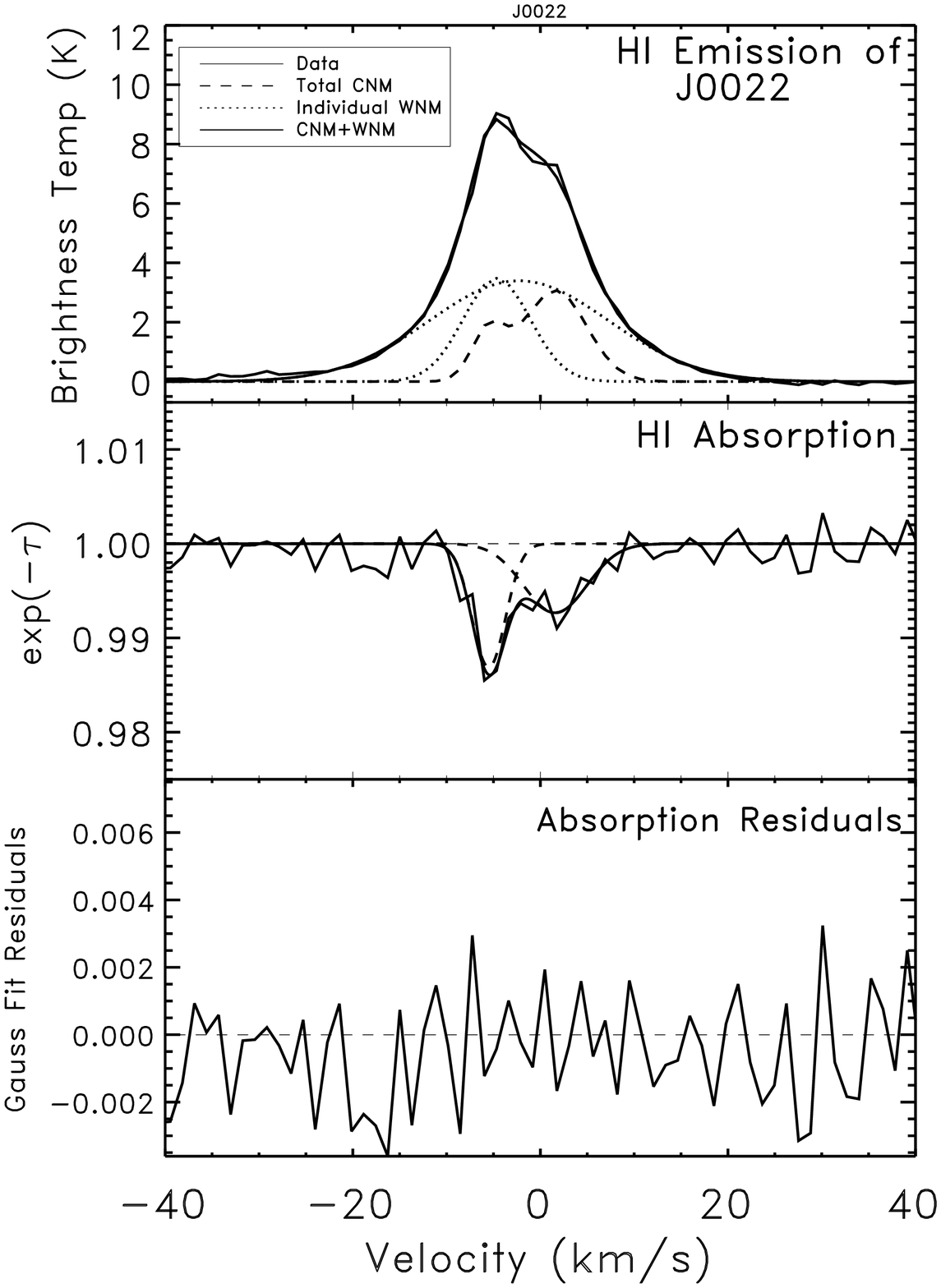}{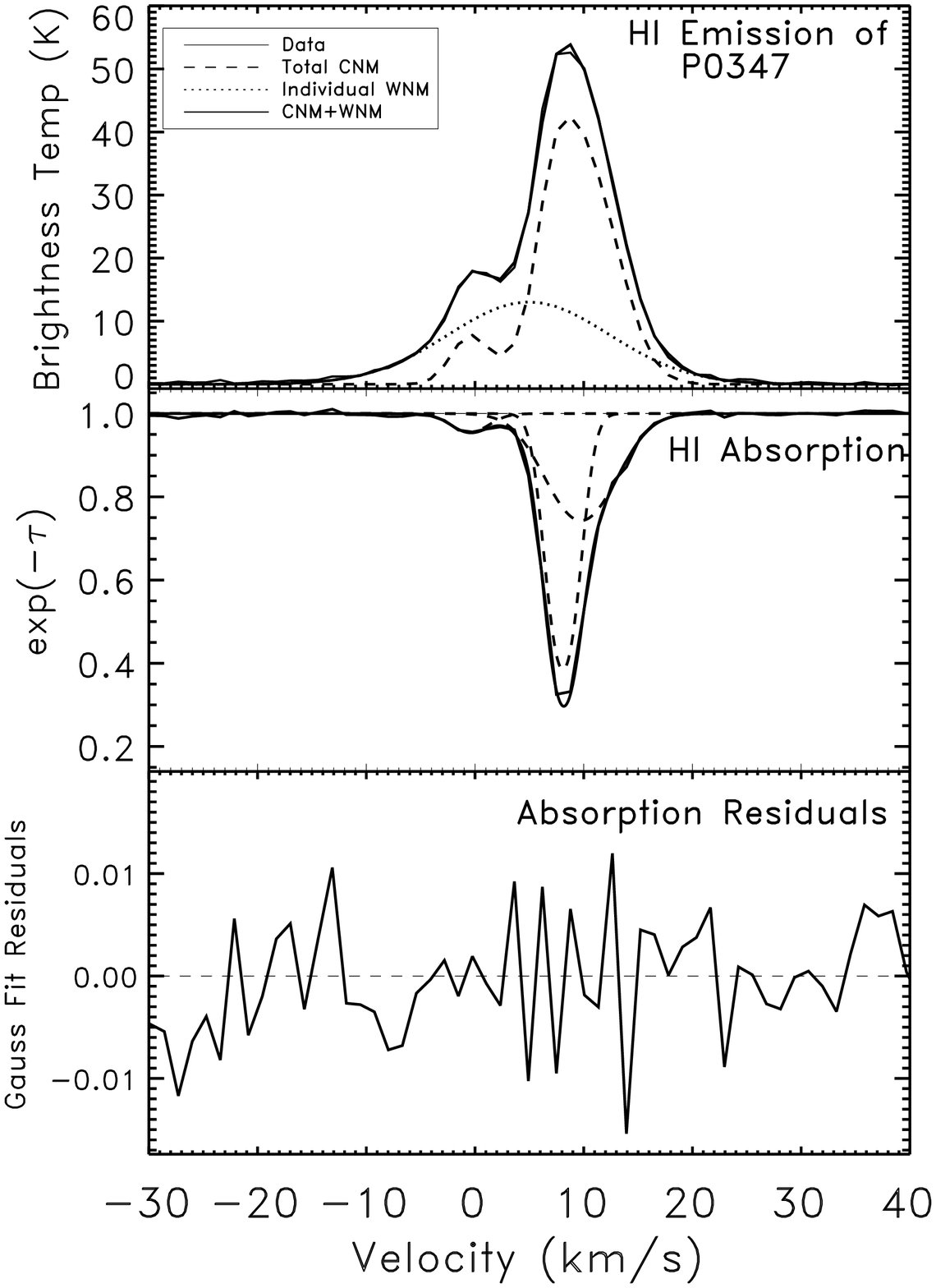}
\caption{\label{f:j0022}
The HI emission and absorption spectra for J0022+002 ({\it Left}) and P0347+05 ({\it Right}). 
The top panel 
shows the Arecibo emission spectrum. The separate contributions from 
the CNM and WNM to the HI emission profile are shown with dashed and 
dotted lines, respectively, while the final, simultaneous fit is 
shown with the thick solid line. The middle panels shows the EVLA 
HI absorption spectrum. The fitted Gaussian functions are shown as 
dashed lines. The bottom panel shows the residuals of  the absorption 
profile after the fitting procedure. 
}
\end{figure*}

We have observed a total of 12 sources and detected HI absorption in 
five directions: 3C286, J0022+002, 3C263.1, P0347+05, and 3C273. 
This results in a detection rate of $\sim 42 \%$ for our experiment. Several detected
absorption features have a peak optical depth of only few $\times 10^{-3}$. 
Therefore, our pilot studies successfully demonstrate the suitability of the EVLA 
for sensitive absorption measurements.

Figures~\ref{f:j0022}$-$\ref{f:3c286} show HI emission and
 absorption spectra for the sources with detected HI absorption.
We have applied the method of HT03 to estimate the spin temperature and the HI column density 
of the absorbing CNM, assuming that the CNM contributes 
to both HI absorption and emission spectra while the WNM contributes only to the HI 
emission spectrum. 
The technique is based on the Gaussian decomposition of both absorption 
and emission spectra, and it takes into account the fact that a certain fraction of the 
WNM gas may be located in front of the CNM clouds, resulting in only a portion of the 
WNM being absorbed by the CNM.

The results of our Gaussian decomposition are over-plotted 
in Figures~\ref{f:j0022}$-$\ref{f:3c286}. 
Three panels are shown for each source. The top panel
shows the Arecibo HI emission spectrum and the separate contributions from the CNM and
WNM to the HI brightness temperature are shown with different lines; the final
(simultaneous) fit to the spectrum is also overlayed. In the middle panel the EVLA HI
absorption spectrum fitted with individual Gaussian components is shown. In the bottom panel
we show the residuals of the absorption spectrum after the fitting process. For each
source the residual spectrum looks random and demonstrates that the fit is perfect 
at the present level of the signal-to-noise ratio and that the Gaussian representation 
of CNM components is valid. The residual spectrum for each source was smoothed to 
various coarser velocity resolutions, up to $\sim$ 25.0 kms$^{-1}$, to search for 
additional broad absorption; no new components were detected.

Table~\ref{tab:cnm} shows the  best fitted CNM parameters. Column (1) gives the source name,
column (2) the peak optical depth ($\tau_{\rm peak}$) 
of the CNM component along with the errorbar, 
column (3) the LSR velocity of the component in km s$^{-1}$, 
column (4) the full width at half maximum (FWHM)
of the absorption component in km s$^{-1}$, and column (5) 
the spin temperature $T_{\rm spin}$ along with its errorbar, in Kelvin. 
In column (6) we list $T_{k,max}$, 
the maximum kinetic temperature in the case of no non-thermal broadening. 
Column (7) gives the HI column density associated with the CNM component along with its errorbar,
in units of 10$^{20}$ cm$^{-2}$, and column (8) 
the upper limit on the CNM HI column density, N(HI)$^{upper}_{CNM}$, also in units of 
10$^{20}$ cm$^{-2}$. To derive N(HI)$^{upper}_{CNM}$ we used used $T_{k,max}$ instead 
of $T_{\rm spin}$.

Table~\ref{tab:wnm} shows the best fit WNM components for each source. 
Column (1) gives the source name,
column (2) the LSR velocity of the WNM component in km s$^{-1}$, column (3) the FWHM in 
km s$^{-1}$, and column (4) the maximum kinetic temperature $T_{k,max}$ in Kelvin.
We discuss the individual sources with HI absorption detections in the following sections.

For sources with no HI absorption detection, we estimated 
limits on the CNM column density and the CNM fraction along each line of sight.
The CNM fraction is defined as the ratio of the CNM column density to
the total HI column density  along the line of sight. Table 4 shows the
3$\sigma$ limit on the peak optical depth, the CNM column density, and
the CNM fraction for sources without a detection of HI absorption.

\subsection{J0022+002 (Figures~\ref{f:j0022}, left)}

No HI absorption was detected in the direction of J0022+002 by \cite{mohan},
whereas our more sensitive EVLA observations detected two absorption
features.  One of the absorption features is weak and wide with 
$\tau_{\rm peak} \sim 7 \times 10^{-3}$ and FWHM $\sim$ 8 km s$^{-1}$.
While T$_{k,max}\sim 1350$ K, our estimated T$_{\rm spin}\sim$ 420 K. 
This component is likely to trace the WNM seen in absorption. However, 
its temperature of 420 K is significantly lower than 
the expected 5000-8000 K and corresponds to the thermally unstable gas.

By fitting Gaussian functions to the emission profile alone, \cite{mohan} found 5 
emission components, with two components; having excitation temperatures 
corresponding to the thermally unstable regime. Our best solution results 
in two emission components, with one component having $T_k <1500$ K and 
corresponding to
the thermally unstable WNM. The total CNM column density along the line of
sight of J0022+002 is $\sim 6 \times 10^{19}$ cm$^{-2}$, 
whereas the total WNM column density
is $\sim $ 2$\times 10^{20}$ cm$^{-2}$; i.e  the total CNM fraction is $\sim$30 \%.

\subsection{P0347+05 (Figures~\ref{f:j0022}, right)}

In addition to the main absorption component at a velocity of $\sim8$ km s$^{-1}$, 
HT03 found a narrow emission feature around V$_{\rm LSR} \sim 0$ km s$^{-1}$ 
without corresponding absorption. This resulted in a thermally 
unstable WNM component. 
Figure~\ref{f:j0022} shows the EVLA absorption
spectrum for this source: our detection of an additional absorption feature at
$-$0.5 km s$^{-1}$ results in the best-fit solution without any thermally-unstable
WNM. 
As discussed in the introduction, this clearly demonstrates that weak absorption lines
can have a significant effect on the estimated fraction of the thermally unstable WNM.
One of the detected absorption features at V$_{\rm LSR}\sim$ 10 kms$^{-1}$ is 
wide and results in T$_{k,max}\sim$1155 K, but with $T_{\rm spin} \sim$ 150 K. 
From Table~\ref{tab:cnm}\&\ref{tab:wnm},
the total CNM fraction along the line of sight of P0347+05 is 69\%.

\begin{figure*}
\epsscale{2.0}
\plottwo{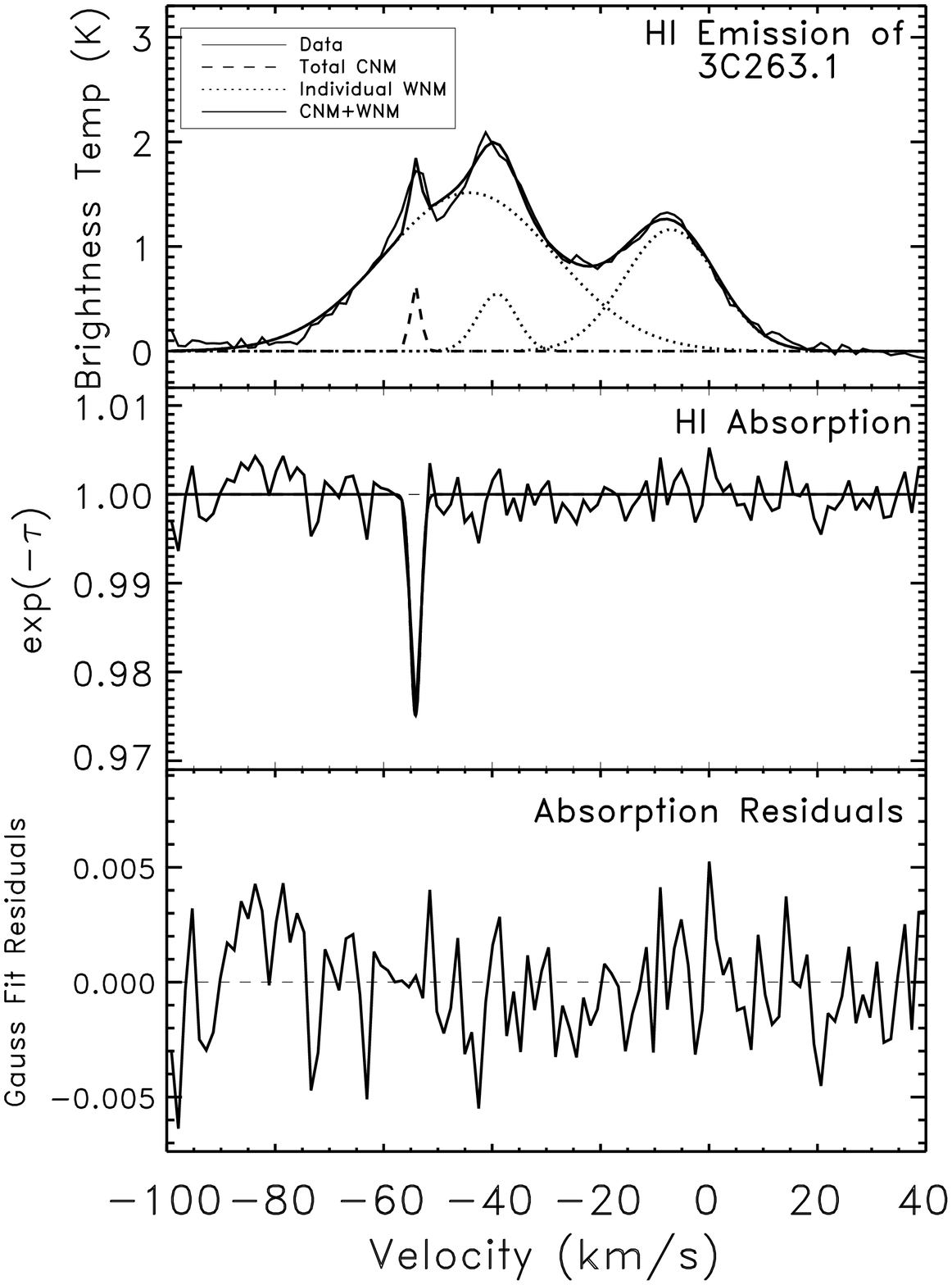}{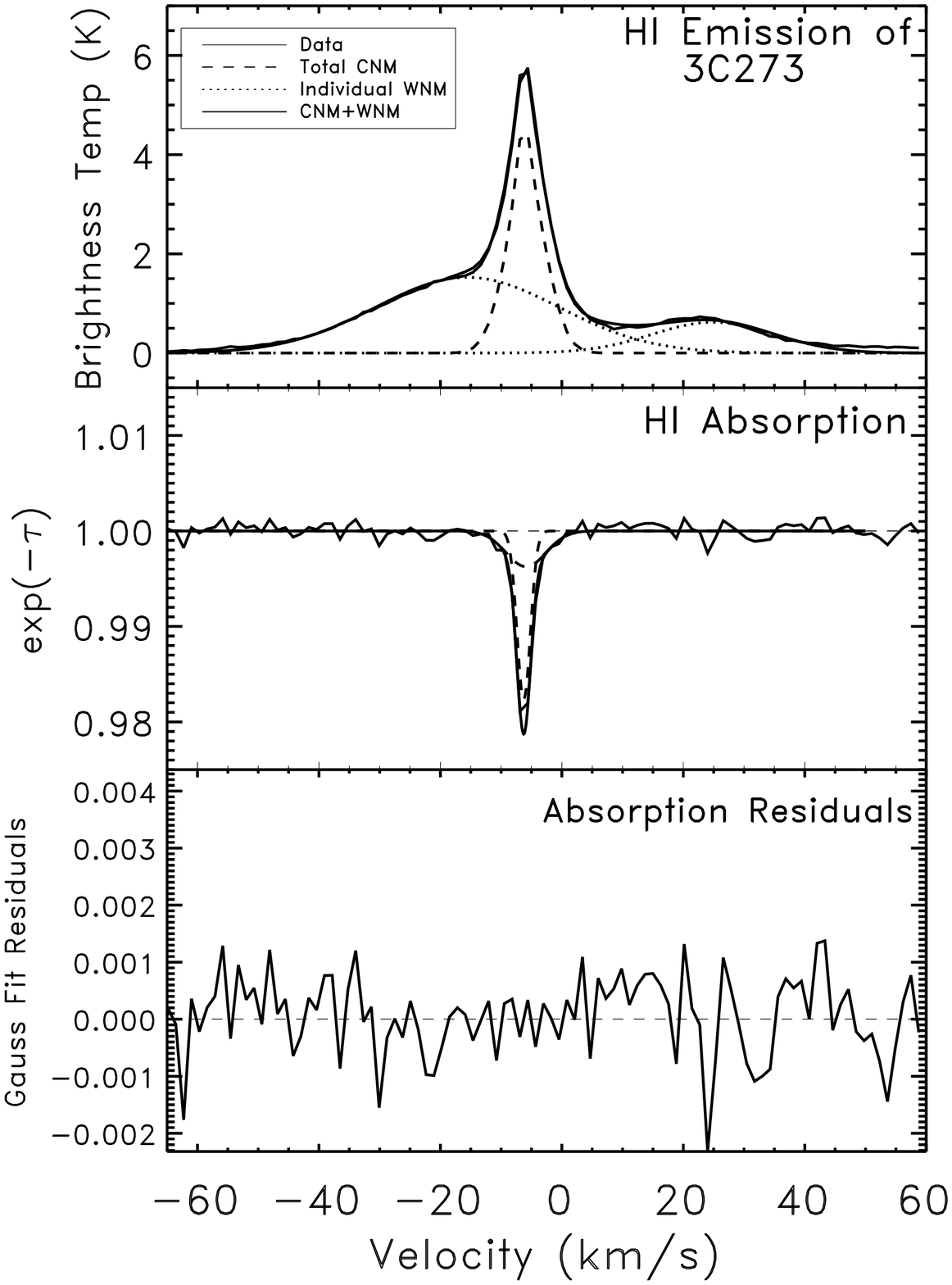}
\caption{\label{f:3c263.1} The HI emission and absorption spectra for 3C263.1 ({\it Left})
and 3C273 ({\it Right}).
The top panel
shows the Arecibo emission spectrum. The separate contributions from
the CNM and WNM to the HI emission profile are shown with dashed and
dotted lines, respectively, while the final, simultaneous fit is    
shown with the thick solid line. The middle panels shows the EVLA
HI absorption spectrum. The fitted Gaussian functions are shown as
dashed lines. The bottom panel shows the residuals of  the absorption
profile after the fitting procedure.
}
\end{figure*}

\subsection{3C263.1 (Figure~\ref {f:3c263.1}, left)}

We detected a narrow HI absorption component towards 3C263.1 at 
V$_{\rm LSR}\sim-$54.0 kms$^{-1}$. 
HT03 did not detect any HI absorption in the direction of 
this source; their best fit to the 
emission/absorption profiles resulted in four emission
components, with three components having T$_{k,max}$ corresponding to 
the thermally unstable WNM.  From our simultaneous fitting of emission and 
absorption profiles, we find that the narrow emission feature 
corresponding to the HI absorption can be reproduced by the CNM alone 
without the need for a thermally-unstable WNM component. 
Again, the detection of weak absorption lines seems to  
decrease the fraction of the thermally unstable WNM. 
Our best-fit solution results in only one thermally unstable WNM component
with T$_{k,max}<1500$ K, while HT03 had three.

The spin temperature for the absorption component  for 3C263.1 is only $\sim30$ K
and is not well constrained by our method. The resulting CNM HI column density,
proportional to this temperature,  is  very low i.e $3 \times 10^{18}$ 
cm$^{-2}$. The total WNM column density in this direction
is 1.6 $\times 10^{20}$ cm$^{-2}$, meaning that the CNM comprises only $\sim$ 2 \% of total
HI column density.

\subsection{3C273 (Figure~\ref {f:3c263.1}, right)}

\cite{dwaraka02} obtained very deep observations with the Westerbork radio telescope (WSRT)
of 3C273 and detected four HI absorption components. 
We detected only two absorption components.
The two addition components lie at the edge of our bandpass, hence they 
were not detected by the EVLA due to the lack of  sensitivity. The two 
absorption components detected by the EVLA are in good agreement 
with \cite{dwaraka02}.

Figure~\ref{f:3c263.1} shows our emission/absorption profiles in the direction of 3C273.
The HI emission profile shows three features; the HI absorption only
corresponds to the narrowest emission component. Using the HT03 method 
we find that the narrow emission feature can be reproduced by the CNM  alone with
a temperature of $\sim70$ K. 
The HI in absorption has a narrow and a wide component.
The wide component has a  FWHM $\sim$ 8 kms$^{-1}$
and $\tau_{\rm peak} \sim 3 \times 10^{-3}$. The corresponding $T_{\rm spin}=900$ K 
and suggests a thermally-unstable WNM. 
From Table~\ref{tab:cnm} \& \ref{tab:wnm}, the total CNM column density 
along the line of sight of 3C273 is $\sim$30 \%.
No HI absorption is detected corresponding to the two additional emission peaks.
This places the 3$\sigma$ upper limit on their spin temperature of 1665 K and 900 K, respectively.

\begin{figure*}
\epsscale{1.0}
\plotone{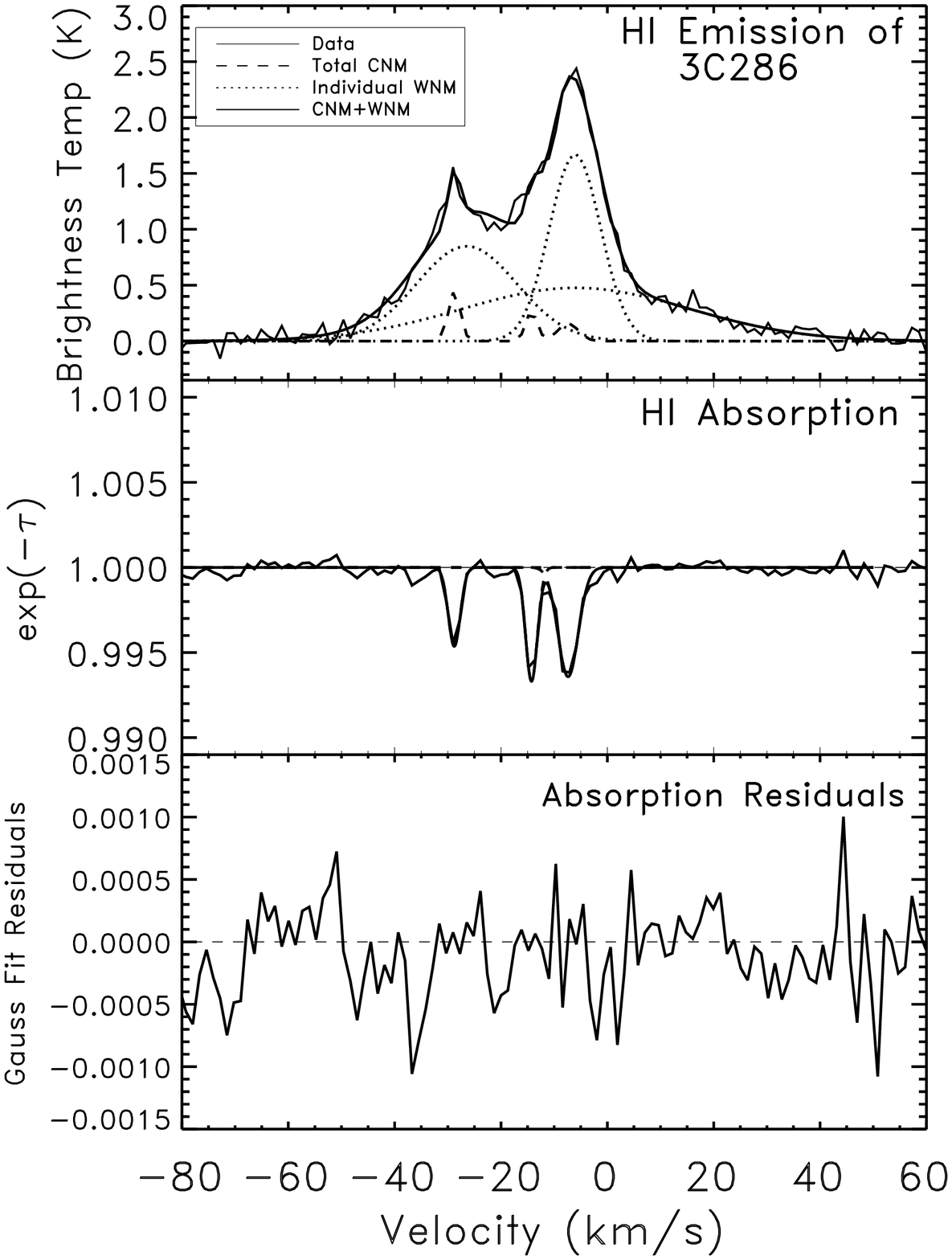}
\caption{\label{f:3c286}The emission and absorption spectra for 3C286.
The top panel
shows the Arecibo emission spectrum. The separate contributions from
the CNM and WNM to the HI emission profile are shown with dashed and
dotted lines, respectively, while the final, simultaneous fit is    
shown with the thick solid line. The middle panels shows the EVLA
HI absorption spectrum. The fitted Gaussian functions are shown as
dashed lines. The bottom panel shows the residuals of  the absorption
profile after the fitting procedure.
}
\end{figure*}

\subsection{3C286 (Figure~\ref{f:3c286})}

HT03 detected two emission components in the direction of 3C286 but no HI absorption.
\cite{sh05} using Arecibo and \cite{bk05} using the WSRT 
detected three weak HI absorption features through much longer integration times. 
As shown in Figure~\ref{f:3c286}, the EVLA absorption spectrum shows three 
distinct features in good agreement with the 
Arecibo and WSRT  detections.  All
three absorption features are weak, with $\tau_{\rm peak}=(4-6) \times 10^{-3}$. 
From our best-fit solutions
to the emission and absorption spectra, we find that two out of three absorption components 
have a very small contribution to the HI emission profiles and their  $T_{\rm spin}$ is not
well constrained. 
The spin temperature of these components are likely to be low, $\sim30$ K.
In addition, their HI column densities are also small, a few $\times$  
$10^{18}$ cm$^{-2}$.  We find that the 
total CNM column density in the direction of this source is 5.2$\times 10^{18}$ cm$^{-2}$, 
whereas the total WNM
column density is 1.2$\times 10^{20}$ cm$^{-2}$. The CNM comprises only 4\% of
total HI column density.
Our best-fit solution results in three emission components; 
with only one  having $T_{k,max}<2800$ K and indicating the presence of thermally unstable gas.

\subsection{Effect of stray radiation}

The emission measurements from single dish telescopes could suffer from stray radiation
entering through the side-lobes of the telescope beam. This could result
in weak emission line wings and therefore affect the estimated spin temperature. 
To check the reliability of our temperature estimates, we compared Arecibo emission spectra 
with the corresponding spectra from the LAB (Leiden-Argentine-Bonn) Galactic 
HI survey \citep{lab}, where the stray radiation correction has been carefully done.

We find that the Arecibo emission spectra for  J0022+002, P0347+05 and 
3C263.1 match well with the LAB spectra. This suggests that the Arecibo
data for these sources are not effected by stray radiation and that our estimates
of spin temperatures are robust.  On the other hand, the Arecibo spectra
for 3C273 and 3C286 show deviations from the LAB spectra.  
In the case of 3C273, the  emission feature
at $\sim25$ km s$^{-1}$ is stronger in the Arecibo data compared to the LAB spectrum, whereas 
the Arecibo spectrum for 3C286 shows  a faint velocity 
tail at $\sim20$ km s$^{-1}$ which is not seen in the
LAB stray-radiation corrected data. 
To check how these differences affect our estimate of T$_{\rm spin}$,
we repeated the Gaussian fitting analysis using the LAB emission spectra for 3C273 and 3C286. 
For both sources, we find that the best T$_{\rm spin}$ solutions
derived using the LAB spectrum are consistent with the Arecibo results within the error-bars. 
For 3C273, the number of thermally stable/unstable WNM components is 
consistent with the Arecibo results.  
However,  for 3C286, the best-fit solution using the LAB data
results in three WNM components, with two being in the thermally unstable regime, 
as compared to only one thermally unstable WNM component when using the Arecibo data.

We conclude that majority of our sources are not effected by 
stray radiation and that the spin temperature derivation is robust.

\section{Discussion and conclusions}

\label{s:discuss}

\begin{table*}
\caption{Limits on non detections }
\centering
\label{tab:nodetect}
\begin{tabular}{lccc}
\hline
\hline
Source  &  $\tau_{\rm peak}$   &          N(HI)$^{a}_{\rm CNM}$  &    CNM Fraction$^{b}$     \\
       &             &          (10$^{20}$ cm$^{-2})$ & (\%) \\
\hline
J0025-260 &    $<0.0027$            &      $<0.016$         &     $<1.2$     \\
3C234 &    $<0.0114$              &      $<0.066$        &       $<4.1$   \\
3C236 &    $<0.0045$                &      $<0.026$   &          $<2.2$\\
3C287 &    $<0.0027$                 &      $<0.016$    &        $<1.4$  \\
J1613+324 &    $<0.0033$          &      $<0.019$      &        $<1.6$  \\
J1634+627 &    $<0.0033$             &      $<0.019$      &     $<1.2$     \\
J1635+381 &    $<0.0048$             &      $<0.028$    &       $<1.5$   \\
\hline
\hline
\end{tabular}
\vskip 0.1 in
${}$~$^a$~N(HI)$_{\rm CNM}$  was derived assuming $T_{\rm spin}$=100 K and FWHM=3 km s$^{-1}$. \\
${}$~$^b$~ The total HI column density was provided by HT03 and Mohan et al. (2004). \\
\vskip 0.1 in
\end{table*}

As mentioned in the previous section, we detected  three wide absorption features 
in the direction of J0022+002, P0347+05 and 3C273, with a velocity FWHM of 7-8 km 
s$^{-1}$, and corresponding T$_{k,max}$ of 1100-1400 K. 
For two of these sources, our best-fit solution resulted in $T_{\rm spin}>$ 400 K. 
These temperatures lie
above the temperature range theoretically allowed for the CNM (Wolfire et al 2003) 
and therefore signify the presence of thermally unstable WNM gas 
with a temperature  $\la 5000$ K \citep{Wolfire03}.
An alternative explanation could be that these broad absorption features represent
a blending of several narrow CNM absorption components.
Our current velocity resolution of 1.6 km s$^{-1}$ does not 
allow us to dismiss this possibility.
Future  EVLA observations with the WIDAR correlator will yeild a higher 
velocity resolution and will allow us to address this question.

The majority of detected absorption lines (8 out of 11) 
in this study have a velocity FWHM $<5$ km s$^{-1}$ and 
the corresponding T$_{k,max}<$ 500 K.
While the 3-$\sigma$ sensitivity of our observations was good enough 
to detect a typical WNM ($\sim10^{20}$ cm$^{-2}$, HT03) 
in absorption with T$_{s}\la$3000 K 
and a FWHM $<12$ km s$^{-1}$, only two broad WNM components were detected.  
Both of these detections have a FWHM $\sim8$ km s$^{-1}$ and  T$_{s}<$1000 K. 
This could be an indication that
the WNM with a temperature $<3000$ K is not as abundant as
indicated previously by indirect observations (e.g. HT03). 
In addition, as we clearly demonstrated in the
case of P0347, the detection of weak HI absorption lines can 
decrease the number of thermally unstable  WNM components.
For the five sources with EVLA HI absorption
detections, previous emission/absorption surveys (HT03 and \cite{mohan}) found
seven thermally unstable WNM components. Our best-fit solutions resulted 
in five thermally unstable components (three seen in emission and two in absorption). 
If applied globally, this would decrease the fraction of 
thermally unstable WNM components from HT03's 
$\sim 39\%$ to $\la27 \%$. We note, however,
that our conclusion is based on a small number 
statistics and that larger and more sensitive emission/absorption
surveys are required to determine the fraction of the 
thermally unstable WNM in a statistically
significant manner.





In several cases (e.g. 3C286 and 3C263.1) we  detected  absorption 
lines which have a low optical depth and a narrow line-width which 
result in a low CNM column density N(HI)$_{\rm CNM}< 5 \times 10^{18}$ cm$^{-2}$.
This is significantly lower than the column density of typical CNM clouds, 
N$_{HI} \sim 10^{20}$ cm$^{-2}$ \citep{Dickey78,Heiles03,Dickey03},
and is close to the lowest theoretical expectations for the CNM \citep{MO77}.
In addition, in these directions
the CNM comprises only $2-$4\% of the total HI column density.  
Sources without detected HI absorption also suggest a very low CNM fraction 
(Table~\ref{tab:nodetect}).
This suggests, as also pointed out by Stanimirovic, Heiles, \& Kanekar (2007), that
the low column density CNM population may be quite abundant in the ISM.
While such small column densities and CNM fractions are hard to explain by
thermal equilibrium ISM models, 
they may be illuminating the tail-end of the CNM population produced
by dynamic and turbulent ISM processes.

\section{Summary}

In order to access the feasibility of sensitive HI absorption observations with the EVLA,
we observed a sample of 12 continuum 
sources, primarily along the lines of sight which have indicated 
the existence of the thermally-unstable WNM (T$<$ 5000 K) in 
previous emission/absorption surveys \citep{Heiles03,mohan}.
The results from this pilot study are as follows.

$\bullet$ For all, except three sources in our sample, we could 
achieve the expected optical depth sensitivity of $\sim10^{-3}$. 
The remaining three sources were resolved with the EVLA; yielding 
a much lower peak flux density, hence an optical depth sensitivity 
$\sim 2-3$ times higher than the expected value.

$\bullet$ Out of 12 sources observed with the EVLA, we detected 
HI absorption in the direction of 5 sources; a detection 
rate of $\sim$42 \% for our experiment.  Seven new absorption features, 
previously undetected in the emission/absorption surveys of 
Heiles \& Troland (2003) and \cite{mohan}, were detected in the 
direction of these 5 sources.

$\bullet$ In the case of several sources e.g. 3C273, P0347+05 and J0022+002,
we detected wide HI absorption lines (FWHM $\sim7-8$ km s$^{-1}$)
resulting in T$_{k,max}<1500$ K. These lines likely trace the
thermally unstable WNM seen in absorption.

$\bullet$ Several detected absorption features  have a peak optical 
depth of only a $\sim$ few $\times10^{-3}$ and trace the tail-end of the 
CNM population. As we demonstrated, these weak HI absorption lines 
play an important role in estimating temperature of the WNM and
can
decrease the indirectly suggested fraction of the thermally unstable WNM.

$\bullet$ We have successfully 
demonstrated the suitability of the EVLA for future sensitive absorption measurements.

\begin{acknowledgements}
The National Radio Astronomy Observatory (NRAO) is a facility of the National Science 
Foundation operated under cooperative agreement by Associated Universities, Inc. 
The Arecibo Observatory is part of the National
Astronomy and Ionosphere Center, which is operated by Cornell
University under a cooperative agreement with the National
Science Foundation.
We thank greatly Lou Nigra (UW-Madison) for kindly obtaining the Arecibo HI emission spectrum
for one of our sources. SS acknowledges support by the Research Corporation for Science Advancement.
\end{acknowledgements}



\begin{thebibliography}{17}
\expandafter\ifx\csname natexlab\endcsname\relax\def\natexlab#1{#1}\fi

\bibitem[{Braun \& Kanekar(2005)}]{bk05}
Braun, R., \& Kanekar, N. 2005, A\&A, 436, L53



\bibitem[{Carilli et al. (1998)}]{carilli98}
Carilli, C. L., Dwarakanath, K. S., \& Goss, W. M. 1998, ApJ, 502, 79



\bibitem[{{Dickey} {et~al.}(2003)}]{Dickey03}
Dickey, J. M., McClure-Griffiths, N. M. Gaensler, B. M., \& Green, A. J. 2003, ApJ, 585, 801



\bibitem[{{Dickey} {et~al.}(1978){Dickey}, {Terzian}, \& {Salpeter}}]{Dickey78}
{Dickey}, J.~M., {Terzian}, Y., \& {Salpeter}, E.~E. 1978, \apjs, 36, 77



\bibitem[{Dwarakanath et al. (2002)}]{dwaraka02}
Dwarakanath, K. S., Carilli, C. L., \& Goss, W. M. 2002, ApJ, 567, 940



\bibitem[{Field {et~al.}(1969)Field, Goldsmith, \& Habing}]{Field69}
Field, G.~B., Goldsmith, D.~W., \& Habing, H.~J. 1969, ApJ, 155, L149



\bibitem[{Hennebelle et al. (2007)}]{ham}
Hennebelle, P., Audit, E, \& Miville-Deschenes, M. A. 2007, A\&A, 465, 445



\bibitem[{{Heiles} \& {Troland}(2003)}]{Heiles03}
{Heiles}, C., \& {Troland}, T.~H. 2003, \apjs, 145, 329



\bibitem[{Kanekar et al. (2003)}]{kanekar}
Kanekar, N., Subrahmanyan, R., Chengalur, J. N., \& Safouris, V. 2003, MNRAS, 346, 57L



\bibitem[{ Kalberla et al. (2005)}]{lab}
Kalberla, P. M. W., Burton, W. B., Hartmann, D., Arnal, E. M., 
Bajaja, E., Morras, R., \& Poppel, W. G. L. 2005, A\&A, 440, 775



\bibitem[{ Liszt(2001)}]{liszt}
Liszt, H. 2001, A\&A, 371, 698



\bibitem[{Mac Low et al. (2005)}]{maclow05}
Mac Low, M., Balsara, D. S., Kim, J., \& de Avillez, M. A. 2005, ApJ, 626, 864







\bibitem[{McKee \& Ostriker(1977)}]{MO77}
McKee, C.~F. \& Ostriker, J.~P. 1977, ApJ, 218, 148



\bibitem[{Mohan et al. (2004)}]{mohan}
Mohan, R., Dwarakanath, K. S. \& Srinivasan, G., 2004, JApA, 25, 143



\bibitem[{Stanimirovi{\' c} \& Heiles (2005)}]{sh05}
{Stanimirovi{\' c}}, S. \& {Heiles}, C., 2005, \apj, 631, 371



\bibitem[Stanimirovi{\'c} et al.(2007)]{2007ASPC..365...22S} 
Stanimirovi{\'c}, S., Heiles, C., 
\& Kanekar, N.\ 2007, SINS - Small Ionized and 
Neutral Structures in the Diffuse Interstellar Medium, 365, 22 



\bibitem[{{Wolfire} {et~al.}(2003){Wolfire}, {McKee}, {Hollenbach}, \&
  {Tielens}}]{Wolfire03}
{Wolfire}, M.~G., {McKee}, C.~F., {Hollenbach}, D., \& {Tielens}, A.~G.~G.~M.
  2003, \apj, 587, 278



\end{thebibliography}

\label{lastpage}

\end{document}